\documentclass[aps,prl,twoside, twocolumn,showpacs,superscriptaddress,groupedaddress]{revtex4}  
\usepackage{graphicx}  
\usepackage{dcolumn}   
\usepackage{bm}        
\usepackage{amssymb}   

\begin{document}

\title{Visualization of the normal-fluid turbulence in counterflowing superfluid $^{4}$He}
\affiliation{National High Magnetic Field Laboratory, 1800 East Paul Dirac Drive, Tallahassee, FL 32310, USA}
\affiliation{Mechanical Engineering Department, Florida State University, Tallahassee, FL 32310, USA}
\affiliation{Department of Physics, University of Florida, Gainesville, FL 32611, USA}
\affiliation{Department of Physics, Yale University, New Haven, CT 06515, USA}
\affiliation{School of Physics and Astronomy, University of Birmingham, Birmingham B15 2TT, United Kingdom}

\author{A. Marakov}
\affiliation{National High Magnetic Field Laboratory, 1800 East Paul Dirac Drive, Tallahassee, FL 32310, USA}
\affiliation{Department of Physics, University of Florida, Gainesville, FL 32611, USA}

\author{J. Gao}
\affiliation{National High Magnetic Field Laboratory, 1800 East Paul Dirac Drive, Tallahassee, FL 32310, USA}
\affiliation{Mechanical Engineering Department, Florida State University, Tallahassee, FL 32310, USA}

\author{W. Guo \footnote{Corresponding: wguo@magnet.fsu.edu}}
\affiliation{National High Magnetic Field Laboratory, 1800 East Paul Dirac Drive, Tallahassee, FL 32310, USA}
\affiliation{Mechanical Engineering Department, Florida State University, Tallahassee, FL 32310, USA}

\author{S.W. Van Sciver}
\affiliation{National High Magnetic Field Laboratory, 1800 East Paul Dirac Drive, Tallahassee, FL 32310, USA}
\affiliation{Mechanical Engineering Department, Florida State University, Tallahassee, FL 32310, USA}

\author{G.G. Ihas}
\affiliation{Department of Physics, University of Florida, Gainesville, FL 32611, USA}

\author{D.N. McKinsey}
\affiliation{Department of Physics, Yale University, New Haven, CT 06515, USA}

\author{W.F. Vinen}
\affiliation{School of Physics and Astronomy, University of Birmingham, Birmingham B15 2TT, United Kingdom}

\date{\today}

\begin{abstract}
We describe a new technique, using thin lines of triplet-state He$_{2}^{*}$ molecular tracers created by femtosecond-laser field-ionization of helium atoms, for visualizing the flow of the normal fluid in superfluid $^{4}$He, together with its application to thermal counterflow in a channel. We show that, at relatively small velocities, where the superfluid is already turbulent, the flow of the normal fluid remains laminar, but with a distorted velocity profile, while at a higher velocity there is a transition to turbulence. The form of the structure function in this turbulent state differs significantly from that found in types of conventional turbulence. This visualization technique also promises to be applicable to other fluid dynamical problems involving cryogenic helium.
\end{abstract}

 \maketitle

In this letter we report two developments relating to the visualization of flow in liquid helium: first a new and powerful technique, based on the observed motion and distortion of a thin line of He$_{2}^{*}$ molecules produced in the liquid by field ionization with a focused femtosecond laser beam; and secondly, the application of this technique in studying the motion of the normal fluid in counterflowing superfluid helium in a more detailed manner than previously possible. The technique promises to be applicable to a wide variety of other fluid dynamical problems involving fluid helium.

Superfluid $^{4}$He exhibits two-fluid behavior~\cite{Tilley1986}: a normal fluid, carrying the thermal energy, coexisting with a superfluid component. Heat transfer occurs by thermal counterflow, the superfluid flowing towards the source of heat and normal fluid away from it. The velocity, $v_{n}$, of the normal fluid is related to the heat flux, $q$, by~\cite{Landau1987}
\begin{equation}
q={\rho}sTv_{n}
\label{eqn1}
\end{equation}
where $\rho$ is the total density and $s$ is the entropy per unit mass. Except at very small heat currents, the effective thermal conductivity of the helium is limited by a force of  mutual friction between the two fluids, and it has been known for many years that this force is due to an interaction (mutual friction) between the normal fluid and a disordered tangle of quantized vortex filaments in the superfluid component~\cite{Vinen1957}. This tangle constitutes a form of turbulence in the superfluid component, usually called quantum turbulence ~\cite{Vinen2002,Skrbek2012}. That quantum turbulence can be generated by the relative motion of the two fluids has been understood for some years from simulations that assume that the flow of the normal fluid remains laminar and spatially uniform~\cite{Schwarz1988,Adachi2010}. However, since thermal counterflow takes place usually in a channel, at the walls of which the normal-fluid velocity must vanish, the assumption that the normal fluid velocity is uniform cannot be correct~\cite{Aarts1994,Baggaley2013}. Furthermore, because the appropriate Reynolds number is typically quite large,  there is the possibility that there is a transition to turbulence in the normal fluid. Indeed, it has been suggested that observed transitions at which the vortex line density suddenly increases are associated with such a transition~\cite{Baggaley2013, Martin1983, Melotte1998}.

The character of the normal fluid flow, in this situation and in many other interesting cases (such as grid turbulence and pipe flow turbulence), is hard to determine except by visualization of the flow. Past experiments on the visualization of thermal counterflow have used micron-sized tracer particles formed from polymer spheres or solid hydrogen~\cite{Sciver2007,Sciver2005,Bewley2006,Paoletti2008A,Mantia2013}. They have produced evidence that flow of the normal fluid is not turbulent at very small heat fluxes, but interpretation has proved difficult at larger heat fluxes because of irregular trapping and de-trapping of the particles on vortex lines~\cite{Kivotides2008}. Here we describe a new technique, using thin He$_{2}^{*}$ molecular tracer-lines created by femtosecond-laser field-ionization, which avoids this and other difficulties (see below) and allows us to obtain detailed information about the normal-fluid flow in both laminar and turbulent regimes.

Our technique is based on our recent demonstration that triplet-state He$_{2}^{*}$ molecules ($a^{3}\sum_{u}^{+}$), with a radiative lifetime in liquid helium of about 13 s ~\cite{McKinsey1999}, can be imaged using laser-induced-fluorescence~\cite{McKinsey2005,Rellergert2008,Guo2009,Guo2010}. These molecules can serve as tracers. Owing to their small size ($\sim$1~nm), such tracers become trapped on vortices only below 0.2~K~\cite{Zmeev2013}. Above 1~K the molecules interact strongly with the normal fluid excitations, so that, like $^{3}$He atoms, they form part of the normal fluid and can be used to trace its motion. The molecules are scattered by the vortices, but this scattering simply contributes to the total mutual friction. In a recent experiment~\cite{Guo2010}, we filled a counterflow channel with He$_{2}^{*}$ tracers, and then tagged a thin line of these tracers by pumping the molecules to the first excited vibrational level with a focussed pump laser pulse. Owing to a low (4\%) tagging efficiency~\cite{Rellergert2008}, acceptable images of this line required averaging over many realizations of the turbulence. A rapid growth of the averaged line-width was observed at higher heat currents, supporting the idea that the normal fluid can become turbulent~\cite{Guo2010}. However, the averaging means the loss of much detailed information.
We have therefore developed a new technique in which a high density of He$_{2}^{*}$ molecules, confined to a line, is created, through field-ionization, by well-focussed femtosecond laser pulses, so that single-shot imaging becomes possible~\cite{Guo2014PNAS}.
%

\begin{figure}[htb]
\includegraphics[scale=0.38]{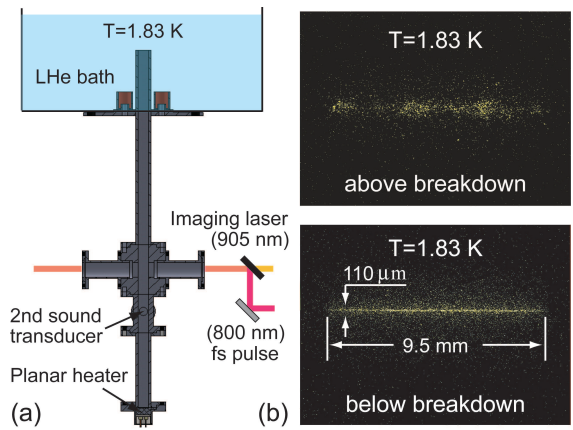}
\caption{(color online). (a) Schematic diagram of the experimental setup. A stainless steel counterflow channel (square cross-section: 9.5~mm$^{2}$; length 300~mm) is attached to a pumped helium bath, the temperature of which is controlled to be 1.83~K for the whole experiment. A planar heater at the lower end of the channel drives the counterflow. The femtosecond laser at 800~nm for tracer-line creation and the imaging laser at 905~nm for tracer-line visualization are combined to pass through a pair of indium-sealed sapphire windows on the channel. An intensified CCD camera views the tracer lines from a front window perpendicular to both the laser beams. (b) Fluorescence images of the He$_{2}^{*}$ molecular tracers created via femtosecond-laser field-ionization in liquid helium. With femtosecond-laser pulse energy above about 60~$\mu$J, dielectric breakdown occurs in helium and isolated clusters of He$_{2}^{*}$ tracers are produced. Slightly below the breakdown pulse energy, a thin line of He$_{2}^{*}$ tracers across the full width of the channel can be produced.} \label{Fig1}
\end{figure}

A schematic diagram of the experiment is shown in Fig.~\ref{Fig1}~(a). The creation of He$_{2}^{*}$ molecules in helium through field-ionization requires laser intensities ${\sim}10^{13}$~W/cm$^{2}$, achieved by focussing a 35-femtosecond pulsed laser (repetition rate 5~kHz) into the channel, with a beam waist about 110~$\mu$m in diameter. Typical fluorescence images of the He$_{2}^{*}$ molecules created in liquid helium at 1.83~K are shown in Fig.~\ref{Fig1}~(b). When the femtosecond-laser pulse energy is above about 60~$\mu$J, dielectric breakdown in helium occurs and isolated clusters of He$_{2}^{*}$ tracers are produced. Slightly below the breakdown pulse energy, a thin line of  He$_{2}^{*}$ tracers across the channel can be produced by controlled electron-avalanche ionization~\cite{Benderskii1999}. Typically 10 pulses are sufficient to produce an adequate tracer concentration. A pair of porous-membrane second sound transducers installed in the channel allows a measurement of the attenuation of second sound in the heat current and so a knowledge of the density of vortex lines~\cite{Vinen2002}.

\begin{figure}[htb]
\includegraphics[scale=0.35]{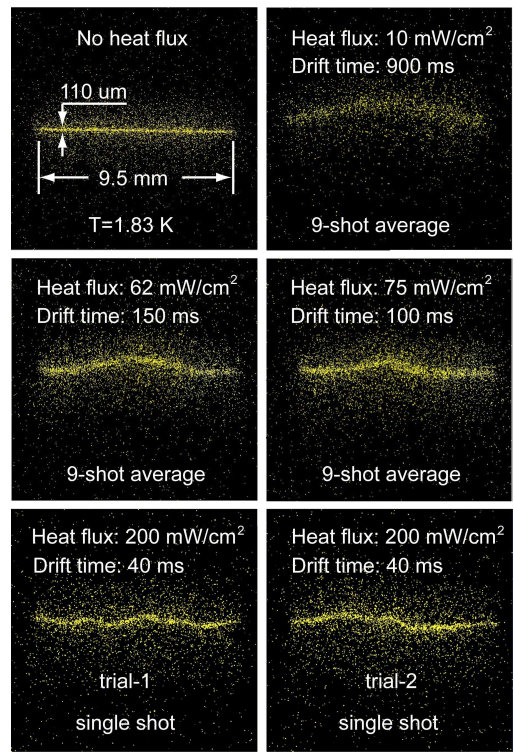}
\caption{(color online). Typical fluorescent images showing the motion of He$_{2}^{*}$ tracer lines in thermal counterflow. At heat fluxes below about 50 mW/cm$^{2}$, an initially straight tracer line always deforms to a nearly parabolic shape, indicating a laminar Poiseuille velocity profile of the normal fluid. As the heat flux is increased to above about 60 mW/cm$^{2}$, the tail part of the tracer line becomes flattened. Above about 80 mW/cm$^{2}$, the tracer line distorts randomly, indicating turbulent flow in the normal fluid. The images shown in the laminar flow regime are averaged over 9 single-shots.} \label{Fig2}
\end{figure}

In a characteristic run, we activate the heater for about 20~s, and then send in femtosecond-laser pulses to create a tracer line. The tracer line moves with the normal fluid for a certain \lq\lq drift time\rq\rq and is then imaged with the 905 nm imaging laser pulses. Fig.~\ref{Fig2} shows typical tracer-line images obtained at various heat fluxes for a temperature of 1.83K (results at 1.60K are very similar). For $q<{\sim}50$ mW/cm$^{2}$, an initially straight tracer line deforms to a nearly parabolic shape, indicating a laminar Poiseuille velocity profile~\cite{Spiga1994} in the normal fluid. For $q>{\sim}80$ mW/cm$^{2}$, we observe random distortions of the lines, indicating large-scale turbulent flow in the normal fluid. For 50~mW/cm$^{2}$ $<q<80$ mW/cm$^{2}$, the outer part of the tracer line becomes flattened. However, for a given $q$ this flattening is reproducible from one set of images to another, so we interpret it as a modification to the laminar normal-fluid profile. Any turbulence on a scale less than about 0.1 mm would lead to anomalous broadening of the line of molecules.  Such broadening is indeed seen for $q>{\sim}80$ mW/cm$^{2}$.  For $q<{\sim}80$ mW/cm$^{2}$ no broadening is seen, confirming that the flow is laminar with no detectable small-scale turbulence.

%

\begin{figure}[htb]
\includegraphics[scale=0.35]{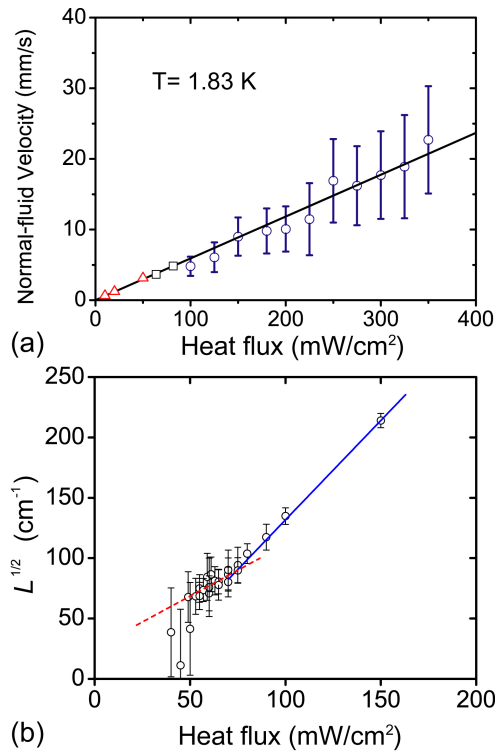}
\caption{(color online). (a) The observed mean normal-fluid velocity $u_{mean}$ as a function of heat flux. The blue circles are data in turbulent flow regime, the black squares are data in the laminar flow regime with tail-flattened velocity profile, and the red triangles are data in the laminar flow regime with near-parabolic velocity profile. The solid line shows the calculated velocity based on Eq.~(\ref{eqn1}) in the text. The error bars are explained in the text. (b) The square root of the measured vortex line density as a function of heat flux in the channel. The solid blue line is a fit to data above $80$~mW/cm$^{2}$, and the dashed red line represent a fit to the data at 50~mW/cm$^{2}<q<80$ mW/cm$^{2}$.} \label{Fig3}
\end{figure}

A local flow velocity, $u$, is obtained by dividing the displacement of a point on an ideally thin tracer line by the drift time. For our tracer lines of finite width, the line image is divided into small segments, and the centre position of each segment is determined with a Gaussian fit of its fluorescence intensity profile. The Lagrangian velocity field obtained in this way involves an average over the drift time and over a volume of order the width of the tracer line. We show in Fig.~\ref{Fig3}~(a) the mean flow velocity, $u_{mean}$, as a function of the heat flux. Mean flow velocity is obtained by averaging the velocity, $u$, over 200 single-shot images and over the length of the line of tracers. The solid line is the velocity given by Eq.~(\ref{eqn1}), indicating that the He$_{2}^{*}$ tracers do follow the normal fluid. In order to investigate the superfluid behavior as the normal fluid undergoes transition to turbulence, we have also measured the vortex line density $L$ using second sound attenuation~\cite{Vinen2002}. The result is shown in Fig.~\ref{Fig3}~(b). Below about 50~mW/cm$^2$, $L$ is too small to be measurable. Above about 80~mW/cm$^{2}$, $L^{1/2}$ grows linearly with the heat flux. The line density coefficient $\gamma$ (defined by $L^{1/2}=\gamma (v_{ns}-v_{0})$, where $v_{ns}$ is the relative velocity of the two fluids) is found to be $162$~s/cm$^{2}$, which is consistent with the reported values found at large heat currents \cite{Martin1983}. At the transition to normal fluid turbulence around $80$~mW/cm$^{2}$, we find no abrupt increase of the vortex line density in disagreement with recent simulations \cite{Baggaley2013}.


\begin{figure}[htb]
\includegraphics[scale=0.35]{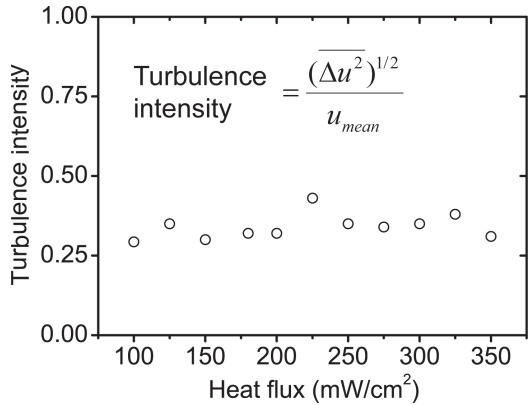}
\caption{(color online). The observed turbulence intensity $\overline{(\Delta{u}^2)}^{1/2}$/$u_{mean}$ of the normal-fluid turbulence as a function of heat flux.} \label{Fig4}
\end{figure}

 In the regime of a turbulent normal fluid, there are large fluctuations in $u$; the error bars in Fig.~\ref{Fig3}~(a) are the averages $\overline{(\Delta{u}^2)}$$^{1/2}$=$\langle \overline{(u-u_{mean})^2}\rangle $$^{1/2}$. The actual probability density function (PDF) for the velocity is found to be close to a Gaussian. The power-law tails observed with particles of hydrogen~\cite{Paoletti2008B} are lost in the present measurements because probing is on too large a length scale~\cite{Mantia2013}. The observed turbulent intensity in the normal fluid, defined as $\overline{(\Delta{u}^2)}^{1/2}$/$u_{mean}$, is shown as a function of heat flux in Fig.~\ref{Fig4}. It is much larger than that observed in classical turbulent channel flow~\cite{Davidson2004}, which is about 5\%. The reason for this intriguing large turbulence intensity is unknown. Note that the turbulent intensity would be smaller if the physically relevant ratio were to involve not $u_{mean}$ but rather $(u_n-u_s)_{mean}$.

\begin{figure}[htb]
\includegraphics[scale=0.4]{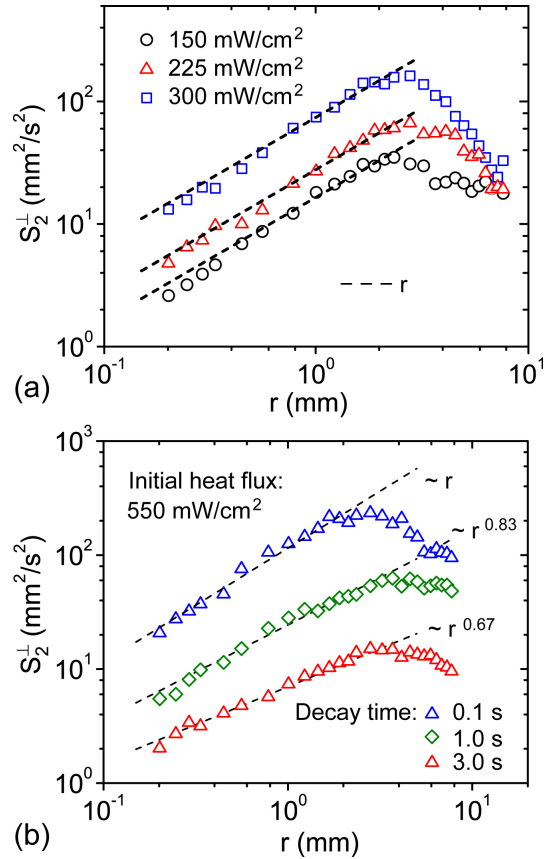}
\caption{(color online) (a) The observed second order transverse structure function $S_{2}^{\bot}$(r) of the normal-fluid turbulence in steady-state counterflow at heat fluxes of 150, 225 and 300 mW/cm$^{2}$, respectively. The black dashed lines represent a power-law form of $S_{2}^{\bot}$(r)$\propto r$. (b) $S_{2}^{\bot}$(r) of the normal fluid turbulence in decaying counterflow with an initial heat flux $550$~mW/cm$^{2}$ at decay times of 0.1 s, 1.0 s and 3.0 s, respectively. The dashed lines represent power-law fits to the data.}\label{Fig5}
\end{figure}

We have used our observations to obtain some information about the second-order transverse structure function, $S_{2}^{\bot}(R, r)=\overline{\langle u(R+r)-u(R)\rangle^2}$, for the special case when the displacement $r$ is along a tracer line and therefore normal to the average flow velocity (the turbulence may not be isotropic). Our results reveal no observable dependence of $S_{2}^{\bot}$ on the reference location $R$, suggesting that the turbulence is approximately homogeneous. The dependence on $r$ for several heat fluxes is shown in Fig.~\ref{Fig5}~(a), where we see that, for $r<2$~mm, $S_{2}^{\bot}$(r)${\propto}r^n$, where $n=1\pm0.05$. This contrasts with $n=2/3$ for classical turbulence in an inertial range with a Kolmogorov energy spectrum (the exponent $n$ leads to an energy spectrum $E(k){\sim}k^{-(n+1)}$). This latter spectrum is believed to appear in decaying counterflow turbulence at sufficiently long decay times such that the two fluids can become coupled through mutual friction \cite{Skrbek2003}. Indeed, we have observed that as the heat current is switched off, $S_{2}^{\bot}$(r) does evolve from the initial form of ${\sim}r$ at short decay times to ${\sim}r^{2/3}$ at long decay times. Fig.~\ref{Fig5}~(b) shows the evolution of $S_{2}^{\bot}$(r) with an initial heat flux of $550$~mW/cm$^{2}$. Our data at other initial heat fluxes also show similar decaying behavior.


Our studies of the flow of the normal fluid in thermal counterflow are still at an early stage. Further study will be reported in due course. Although we only present experiments on fluid flow in helium-II, the flow of helium-I and cryogenic gaseous helium also promises to be of great interest. Since its viscosity can be easily tuned \cite{Threlfall1975} and can span orders of magnitude greater than that of commonly used classical fluids (water, air, SF$_{6}$), cryogenic helium can be used in the study of classical flows over large ranges of Reynolds and Rayleigh numbers. Particularly, cryogenic helium has been used in compact ($L \sim 1$~cm) experiments reaching Reynolds numbers up to $\sim 10^{7}$ \cite{Fuzier2001}. A similar experiment using conventional classical fluids would be orders of magnitude greater in both size and expense, requiring extremely specialized facilities \cite{Zargola1998}. Further, since our technique is applicable in \textit{any} phase of helium, it is particularly relevant to the studies of large circulations and structures, such as plumes, which have high Rayleigh (and Reynolds) number~\cite{Donnelly-Sreeni}, but have been hampered by the lack of effective visualization and velocimetry techniques in gaseous helium.

We acknowledge the startup support provided to W.G. by Florida State University and the National High Magnetic Field Lab (NHMFL), as well as support from the US Department of Energy under Grant DE-FG02 96ER40952, the National Science Foundation under Grant No. DMR-1007974 and Grant No. DMR-1007937, and the Engineering and Physical Sciences Research Council in United Kingdom under Grant No. EP/H04762x/1. The authors would also like to acknowledge the machine shop people at the NHMFL and B. Malphurs at University of Florida for their help on the construction and installation of the optical cryostat and the flow channel.

\end{document}